\documentclass{raa}            
\usepackage{graphicx,times}             
\usepackage{threeparttable}
\usepackage[rotateright]{rotating}
\usepackage{lscape}

\input{epsf.sty}                        
\input{psfig.sty}                       

\begin{document}

\title{Interchange reconnection between an active region and a coronal\\ hole \thanks{Supported by the National Science Foundation of China.}}

   \volnopage{Vol.0 (200x) No.0, 000--000}
   \setcounter{page}{1}

\author{Lin Ma\inst{1,2}
    \and Zhong-Quan Qu\inst{1}
    \and Xiao-Li Yan\inst{1}
    \and Zhi-Ke Xue\inst{1,2}}

\institute{National Astronomical Observatories/Yunnan Observatory,
 Chinese Academy of Sciences, Kunming 650011, China; {\itshape{malin-567@ynao.ac.cn}}\\
 \and Graduate University of Chinese Academy of Sciences, Beijing 100049, China}

 \date{Received~~ month day; accepted~~ ~~month day}

\abstract{With the data from the Atmospheric Imaging Assembly (AIA) and the Helioseismic and Magnetic Imager (HMI) on board the Solar Dynamics Observatory (SDO), we present a magnetic interaction between an isolated coronal hole (CH) and an emerging active region (AR). The AR emerged nearby the CH and interacted with it. Bright loops constantly formed between them, which led to a continuous retreat of the CH boundaries (CHBs). Meanwhile, two coronal dimmings respectively appeared at the negative polarity of the AR and the east boundary of the bright loops, and the AR was partly disturbed. Loop eruptions followed by a flare occurred in the AR. The interaction was also accompanied by many jets and an arc-shaped brightening that appeared to be observational signatures of magnetic reconnection at the CHBs. By comparing the observations with the derived coronal magnetic configuration, it is suggested that the interaction between the CH and the AR excellently fitted in with the model of interchange reconnection. It appears that our observations provide obvious evidences for interchange reconnection.
}
\authorrunning{Lin Ma et al}
\titlerunning{interchange reconnection between AR and CH}

\maketitle

\section{Introduction}
\label{sect:intro}
Coronal holes (CHs) appear as dark areas when observed with X-ray and extreme-ultraviolet (EUV) lines (Altschuler et al. 1972; Vaiana et al. 1976), due to their lower densities and lower temperatures (Munro \& Withbroe 1972). Different from active regions (ARs) and quiet sun (QS), CHs are dominated by unipolar magnetic fields in which magnetic field lines are open outwards to interplanetary space. Plasma can be evacuated along the open field lines of CHs, generating high-speed solar wind streams which may induce geomagnetic storms on the earth (Krieger et al. 1973; Fisk \& Schwadron 2001; Tu et al. 2005). According to their locations, CHs can be classified into two types: polar CHs and mid-latitude CHs. The mid-latitude ones can be either ``isolated'' or equatorial extensions of polar CHs (EECHs). In contrast to the typical differential rotation of photosphere, EECHs tend to show quasi-rigid rotation (Timothy et al. 1975; Wang et al. 1996). In such a case, magnetic reconnection is believed to occur at CH boundaries (CHBs) to maintain the CH integrity (Fisk et al. 1999; Fisk \& Schwadron 2001; Kahler \& Hudson 2002; Fisk 2005; Wang \& Sheeley 2004).

In recent years, the process of reconnection at CHBs has been generally accepted as ``interchange reconnection'' (Crooker et al. 2002). Such magnetic reconnection occurs between open filed of CHs and closed field of ARs or QS, and its result is an exchange of footpoints between open and closed field lines, with conservation of total open or closed flux. Wang \& sheeley (2004) presented two kinds of interchange reconnection. The first typically occurs when a magnetic bipole emerges inside or at the boundary of an open field region. An X-point which may be at any height marks the intersection between the open flux and closed flux. Reconnection at the X-point transfers the closed line in the opposite direction, while the footpoint of the open field line undergoes a discontinuous jump. The second is inherently three-dimensional, involving interaction between open field line and neighboring but non-coplanar closed loop. The reconnection occurs at the apex of the closed loop and leads to stepwise displacements within a region of single magnetic polarity. If the closed loop separate two open field regions of opposite polarity, its apex is consistent with the Y-point at the source surface. In this paper, we call them X-type and Y-type interchange reconnection, respectively.

Although interchange reconnection has been involved in a wide field of CHs, such as maintaining the quasi-rigid rotation of CHs and triggering small-scale evolution of the CHBs (Nolte et al. 1978; Madjarska \& Wiegelmann 2009), its observational evidences are rather rare so far. Kahler \& Moses (1990) studied X-ray images of an EECH and found that X-ray bright points at CHBs played an important role in the expansion and contraction of the EECH. Kahler \& Hudson (2002) investigated the boundary morphology of three EECHs, however, no obvious signatures of magnetic reconnection were found. By means of spectroscopic observations, Madjarska et al. (2004) provided for the first time observational evidences for interchange reconnection. The reconnection was proved by bidirectional jets occurring along CHBs. Subsequently, Attrill et al. (2006) gave evidence for interchange reconnection forced by an expanding coronal mass ejection (CME). They argued that the asymmetric evolution of two coronal dimmings caused by the CME eruption represented credible signature of the reconnection. As more direct evidences given by Baker et al. (2007), new bright loops forming between a CH and an emerging AR, retreat of the CHBs, and coronal dimming appearing on one side of the AR jointly indicated the occurrence of interchange reconnection between the CH and the AR. Similar evidences were also reported by Yokoyama \& Masuda (2010).

On 2010 June 7, an emerging AR appeared close to an isolated CH and interacted with it. We expect that the interaction is just interchange reconnection. Therefore, using the EUV images from the Solar Dynamics Observatory (SDO), we examined if there were some observational evidences of interchange reconnection during the interaction. In Section 2 and 3, the data analysis and the results are presented, respectively. The discussions are given in Section 4, while the summary is in Section 5.

\section{DATA ANALYSIS}
\label{sect:Dat}
The data used in the present study were obtained by the Atmospheric Imaging Assembly (AIA; Lemen et al. 2012) and the Helioseismic and Magnetic Imager (HMI; Schou et al. 2012) on board the SDO. The AIA employs seven narrow EUV and three UV-visible-light bandpasses, providing multiple simultaneous high-resolution full-disk images with 1.5 arcsec spatial resolution and 12 s cadence. For this event, we mainly used three EUV channels, respectively centered at wavelengths 193 {\AA}, 171 {\AA}, and 304 {\AA}. Since 193 {\AA} basically reveals the information of the corona, it was chosen as the most appropriate one to determine the CHBs and to study the CH evolution. The HMI observes the full Sun at six different wavelengths and six polarization states in the Fe $\textrm{I}$ 6173 {\AA} absorption line. Here, we used line-of-sight magnetograms with 45 s time cadence and a pixel size of 0.5$^{''}$ to analyze the photospheric magnetic field. To trace the flare time, we also examined the SXR flux observed by the Geostationary Operational Environmental Satellite (GOES).

The AIA images and HMI magnetograms were derotated to a reference time (14:00 UT on 7 June 2010). The CHBs were defined as the regions with intensities 1.5 times the average intensity of the darkest region inside the CH (Madjarska \& Wiegelmann 2009; Yang et al. 2011).

\section{Result}

The interaction occurred between the isolated CH and the emerging AR. It originated from the expansion of the AR and was accompanied by obvious change of the CH. Figure 1 shows the interaction process and associated phenomena in AIA 193 {\AA} images. The CH was centered at about N21$^{\circ}$E22$^{\circ}$, with latitudinal extension of about 12$^{\circ}$ and longitudinal width of nearly 8$^{\circ}$. Nearby, the AR lay to the southeast of the CH. It began to emerge at about 01:00 UT June 7 as a bipole oriented in an east-west direction, and then developed in both magnetic flux and area. Figure 1($a$) indicates that the CH predominantly had negative-polarity magnetic fields in the photosphere and was closer to the leading negative polarity of the AR.
In AIA observations, three remarkable phenomena suggest the occurrence of the interaction. The first is that bright loops formed between the AR and the CH. When the AR was small, it was entirely encircled by short loops (see panel (b)). Then following its expansion, new bright loops started to form towards the CH at about 06:24 UT (indicated by thin white arrow). This kind of loops constantly appeared and gradually developed into an extensive loop system which connected the AR and the CH (see thin white arrows in panels (b2-b8)). By comparison with the HMI magnetogram, it is clear that the new bright loops are linking the positive polarity of the AR and the negative CH field (see panel ($a$)). The second is that the CH changed seasonably as the loop system formed. Taking 06:24 UT as the initial time of the loop formation, the CH was almost unchanged theretofore (see panels (b) and (b1)). However, after that the CHBs persistently receded and the CH area gradually shrunk, which can be easily seen from the white contours in the figures (panels (b1-b8)). It seems that the new bright loops closed down the CH field, leading to the boundary retreat and the area shrinkage of the CH. The last is that two dimming regions, ``D1'' and ``D2'', appeared during the retreat of the CHBs. On either side of the new loop system, they were visible in the 193 {\AA} fixed-base difference images. D1 was a compact region immediately adjacent to the negative polarity of the AR, while D2 showed up as a more diffuse one at the east boundary of the new loop system. We deem that these three phenomena represented creditable manifestations of the interaction. Therefore, it is believed that the interaction indeed occurred and was almost simultaneous with the formation of the new bright loops. During the interaction, the AR was very active. Loop eruptions accompanied by a flare of X-ray class B2.0, with start, peak and end times respectively around 19:15, 19:24, and 19:31 UT, occurred in the AR (see panel (b7)). This activity of the AR may involve two origins, the own development of the AR or the disturbance from the interaction. In other words, it can be considered as an alternative manifestation of the interaction.

In Figure 2, six space-time plots along six slits (``A-B'', ``C-D'', and ``E-F'' in  Figure 1(b6), and ``A$'$-B$'$'', ``C$'$-D$'$'', and ``E$'$-F$'$'' in Figure 1(b7)) are displayed to analyze the formation of the new bright loops and the retreat of the CHBs. We obtained the profiles in the plots by averaging five pixels in 193 {\AA} images in the directions perpendicular to the slits over time from 06:00 UT to 18:00 UT. Linear fitting was used to derive the formation velocity of the new bright loops and the retreat velocity of the CHBs at each site. At the three sites ``A-B'', ``C-D'', and ``E-F'', the formation velocities are respectively 0.914 $\pm$ 0.024 km s$^{-1}$, 1.565 $\pm$ 0.033 km s$^{-1}$, and 1.671 $\pm$ 0.042 km s$^{-1}$. At the three sites ``A$'$-B$'$'', ``C$'$-D$'$'', and ``E$'$-F$'$'', the retreat velocities are respectively 0.368 $\pm$ 0.020 km s$^{-1}$, 0.515 $\pm$ 0.016 km s$^{-1}$, and 0.557 $\pm$ 0.013 km s$^{-1}$.

To examine the interaction in detail and in the round, its zoomed-in view in 193 {\AA}, 171 {\AA}, and 304 {\AA} images is presented in Figure 3.
As a notable characteristic of the interaction, many EUV jets occurred at the CHBs during the formation of the loop system between the AR and the CH. They could be identified by eyes according to their evolution in both 193 {\AA} and 171 {\AA} movies. In this paper, we only show the jets in 193 {\AA} images. Figure 3($a$1-$a$3) are three 193 {\AA} images with superposition of sub-images outlined by white rectangles and timetables. The sub-images exhibit the jets occurring at different times and indicate their sites, while the timetables list the occurrence time of each jet. As shown in the figures, eleven sites with bright jets were observed at the CHBs. The eleven sites included the east boundary, footprints, and interior of the new bright loops. It is noted that the jets displayed in rectangle ``1'', rectangle ``2'', and rectangle ``3'' occurred when the new bright loops whose footprints adjoin them formed.
According to previous studies, jets are transient plasma ejections triggered by magnetic reconnection and are thus believed as observational signatures of magnetic reconnection at the CHBs (Yokoyama \& Shibata 1995; Subramanian et al. 2010; Yang et al. 2011). Therefore, these jets in this study may suggest the occurrence of magnetic reconnection. Another distinct characteristic of the interaction is that an arc-shaped brightening appeared at the CHBs, accompanying the formation of the loop system between the CH and the AR. The brightening can be clearly seen in both 171 {\AA} and 304 {\AA} fixed-base difference images in Figure 3 (indicated by white arrows). Almost along the footprints of the new bright loops, it developed to be an arc-shaped enveloping surface of the loop system. Jiang et al. (2007) indicated that, brightenings formed at CHBs may root in the released energy of magnetic reconnection between a CH and an erupting filament. Similarly, it is probable that the arc-shaped brightening in our study had the same origin.

\section{DISCUSSION}

The interaction inherently occurred between the magnetic field of the AR and that of the CH. It is now known that AR fields are closed fields, while CH fields are dominated by open fields. In view of such knowledge and the observations described above, we surmise the following: the expansion of the AR's closed field drove magnetic reconnection with the CH's open field. Such reconnection led to the formation of closed field lines between them and opening of field lines at the negative polarity of the AR. Then, the closed field lines between the CH and the AR shut down the CH field, making the CHBs be out of shape, and open filed lines accumulated at the AR's negative polarity to form D1. In particular, the EUV jets and the arc-shaped brightening at the CHBs represented creditable signatures of the magnetic reconnection. If this is true, the large-scale coronal configuration should have some reflections. Using the results of the Potential-Field Source-Surface (PFSS) model (Schrijver \& DeRosa 2003) based on the synoptic magnetic maps from the Michelson Doppler Imager (MDI; Scherrer et al. 1995) with a 6 hr time resolution, we construct the coronal configuration. The representative coronal magnetic field lines are shown in Figure 4. There are two kinds of coloured lines in the figure. The yellow lines represent closed filed lines, while the green lines represent open field lines. As the key of the conjecture, the closed field lines linking the positive magnetic field of the AR and the CH field and the open filed lines at the AR's negative polarity clearly exist in this configuration. It appears that the interaction between the AR's closed field and the CH's open field resulted in the formation of new closed and open field lines, which partly accorded with the model of interchange reconnection. Moreover, the magnetic configuration reveals that, the AR field and the CH field are favourable for Y-type interchange reconnection when their field lines are non-coplanar. Therefore, we believe that the conjecture is reasonable and the interaction should be interchange reconnection. In Figure 5, the sketch of Y-type interchange reconnection (as shown in Figure 4 of Wang et al. 2004) is presented to interpret the interaction.

The closed loop of the AR (A) expands and reconnects at its apex with the non-coplanar CH's open field line (B). After reconnection, the closed loop (A) is transported to northwest to form new closed loop (C) which connects the positive polarity of the AR and the negative CH field. The new closed loop (C) is exactly corresponding to the bright loop shown in Figure 1. As well as shifting the closed loop (A), reconnection also transports the open field line (B) to the negative polarity of the AR, creating new open field line (D). Such successive reconnection due to the expansion of the AR will close down the CH field leading to the boundary retreat and the area shrinkage of the CH. Meanwhile, open field lines will accumulate at the AR's negative polarity to form dimming region D1 (shown in Figure 1). Plasma is no longer trapped in the closed AR loops but evacuated along the open field lines of D1. Seen from the coronal configuration, some open field lines in D1 pass through the dimming region D2, so we speculate that D2 is caused by mass deletion. Furthermore, the AR may be more eruptive since the restriction of the overlying closed loops on the AR is partially removed by the reconnection.

Baker et al. (2007) gave observational evidences for interchange reconnection between a CH and an emerging AR. The reconnection was evidenced by bright loops linking the CH and the AR, the retreat of CHBs, and coronal dimming on one side of the AR. In this study, we not only observe these three phenomena but also find the arc-shaped brightening and many EUV jets that have been accepted as observational signatures of magnetic reconnection by many researchers. Moreover, we also observe the AR's disturbance expected by them. Therefore, we deduce that our observations provide obvious evidences for interchange reconnection.

\section{SUMMARY}

We present observational evidences for interchange reconnection between the CH and the AR. We find that bright loops connecting the AR's positive polarity and the CH negative field formed, which led to the boundary retreat and the area shrinkage of the CH. At the same time, two coronal dimmings respectively appeared at the AR's negative polarity and the east boundary of the bright loops, and many jets and the arc-shaped brightening that were typical signatures of magnetic reconnection were observed at the CHBs. Moreover, the AR was partly disturbed. Loop eruptions followed by a flare of class B2.0 occurred in the AR. These phenomena jointly constitute the observational signatures of the interchange reconnection.

It is generally appreciated that interchange reconnection plays an important role in the small-scale and short-time evolution of CHBs. However, our example strongly suggest that interchange reconnection can also cause large-scale evolution of CHBs on short timescales. If it sustains for enough time, the original CH may be disappear and new CH may form. Therefore, interchange reconnection could be regarded as an alternative approach to explain the formation or disappearance of CHs on short timescales.

In this study, the average formation velocity of the bright loops between the CH and the AR is about 1.383 km s$^{-1}$, and the average retreat velocity of the CHBs is about 0.48 km s$^{-1}$. Both of them may be associated with the rate of interchange reconnection. However, it is difficult to know the real rate of the interchange reconnection from the two velocities. So, more data and techniques are needed to quantize interchange reconnection.

\begin{acknowledgements}

We thank the referee for many constructive suggestions that improved
the quality of this paper. We acknowledge the SDO teams for
providing the data. SDO is a mission for NASA's Living with a star
(LWS) program. We also thank the GOES teams for granting free access
to their Internet databases. This work is supported by the National
Science Foundation of China (NSFC) under grant 11078005, 11373066,
11370365.

\end{acknowledgements}

\setcounter{figure}{0}
\begin{figure*}[h]
\centering
\includegraphics[width=16cm]{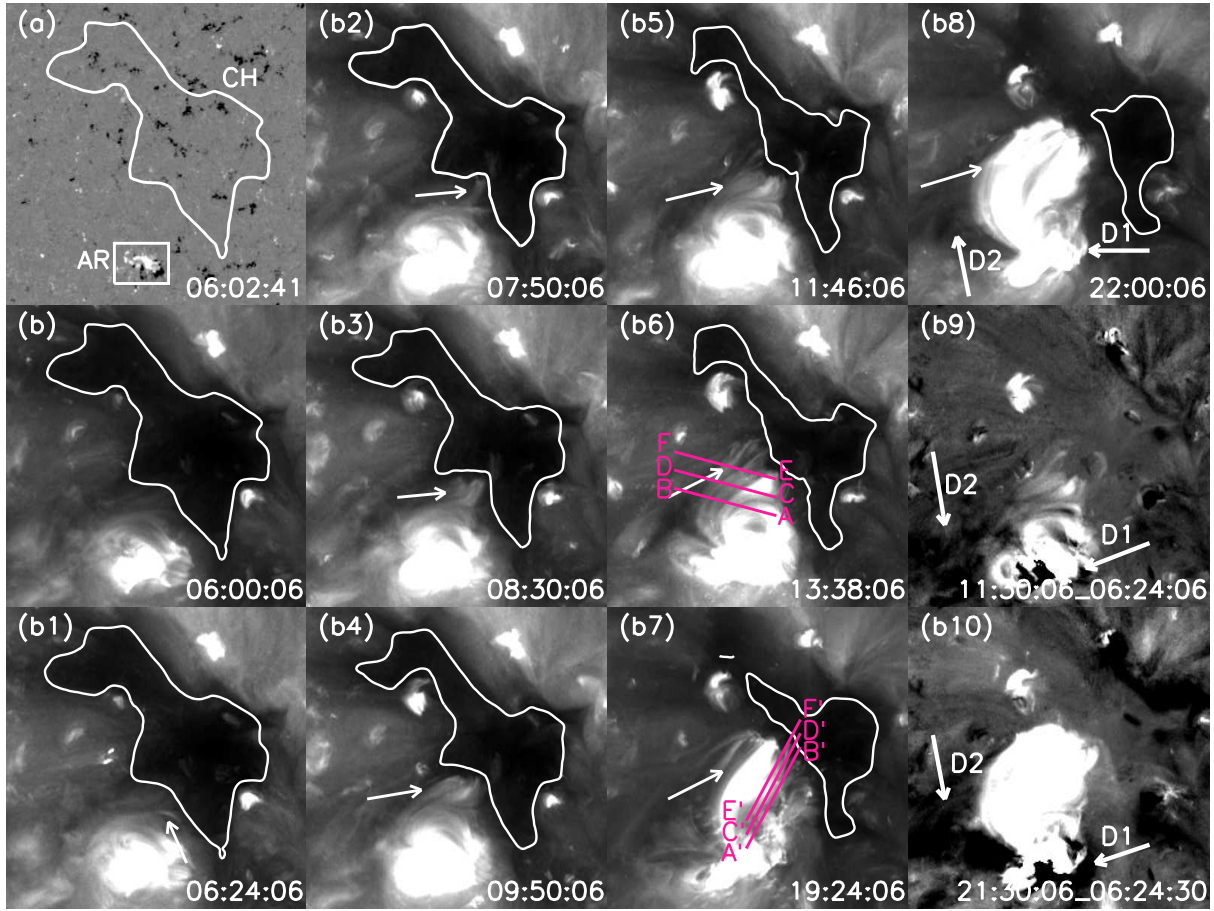}
\caption{HMI magnetogram ($a$) with superposition of outline of the CH determined from panel (b), AIA 193 {\AA} direct images (b-b8), and 193 {\AA} fixed-base difference images (b9-b10), showing the interaction process. The white contours represent the CHBs. The thin white arrows in panels (b1-b8) indicate the new bright loops forming between the AR and the CH. The thick white arrows, ``D1'' and ``D2'', mark the two dimming regions. The magenta slits ``A-B'', ``C-D'', and ``E-F'' in panel (b6), and ``A$'$-B$'$'', ``C$'$-D$'$'', and ``E$'$-F$'$'' in panel (b7) outline the position for analyzing the  temporal variation of 193 {\AA} intensities in Figure 2. The field of view (FOV) is 260$^{''}$ $\times$ 260$^{''}$.}
\end{figure*}

\setcounter{figure}{1}
\begin{figure*}
\centering
\includegraphics[width=7.45cm]{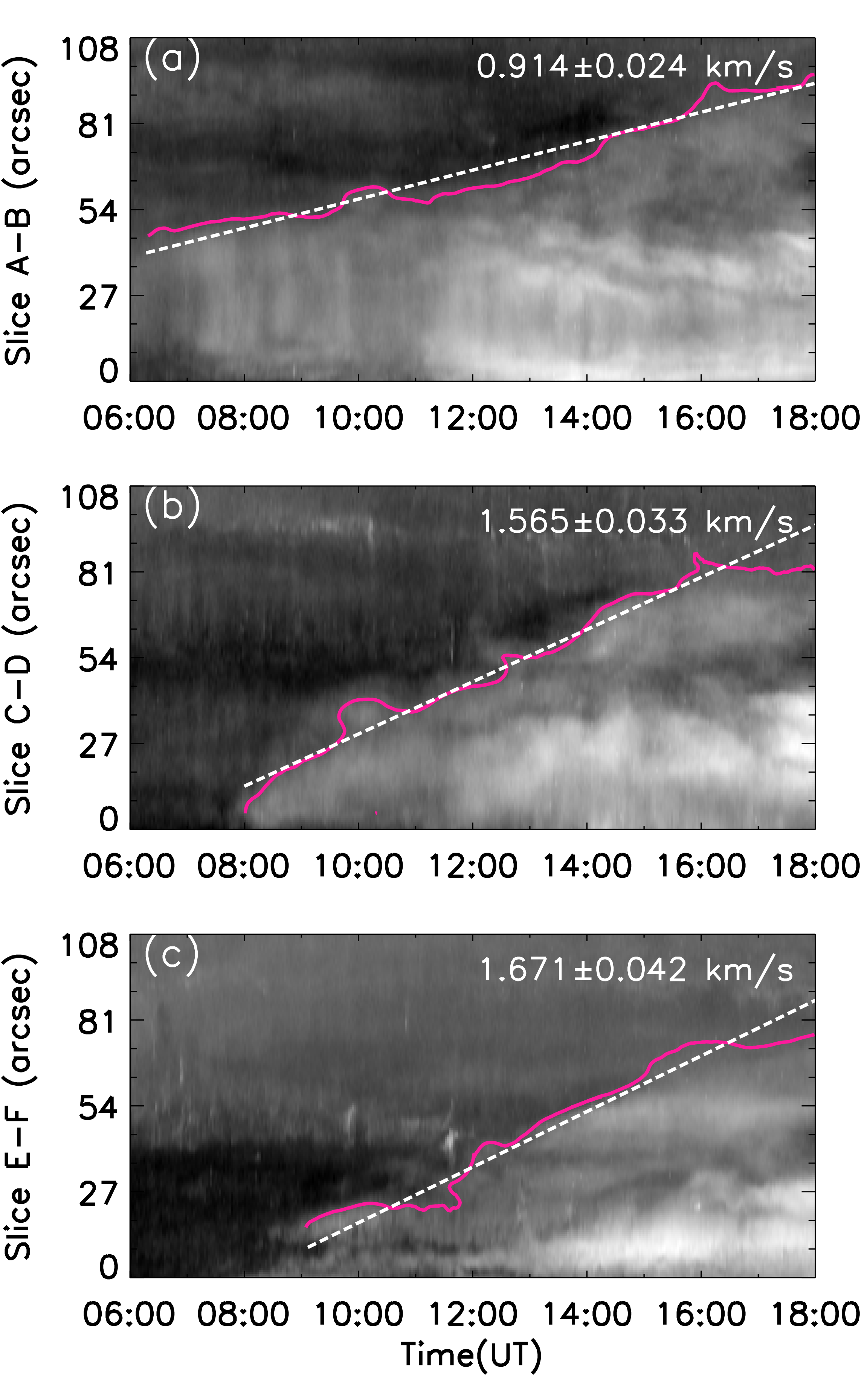}
\includegraphics[width=7.45cm]{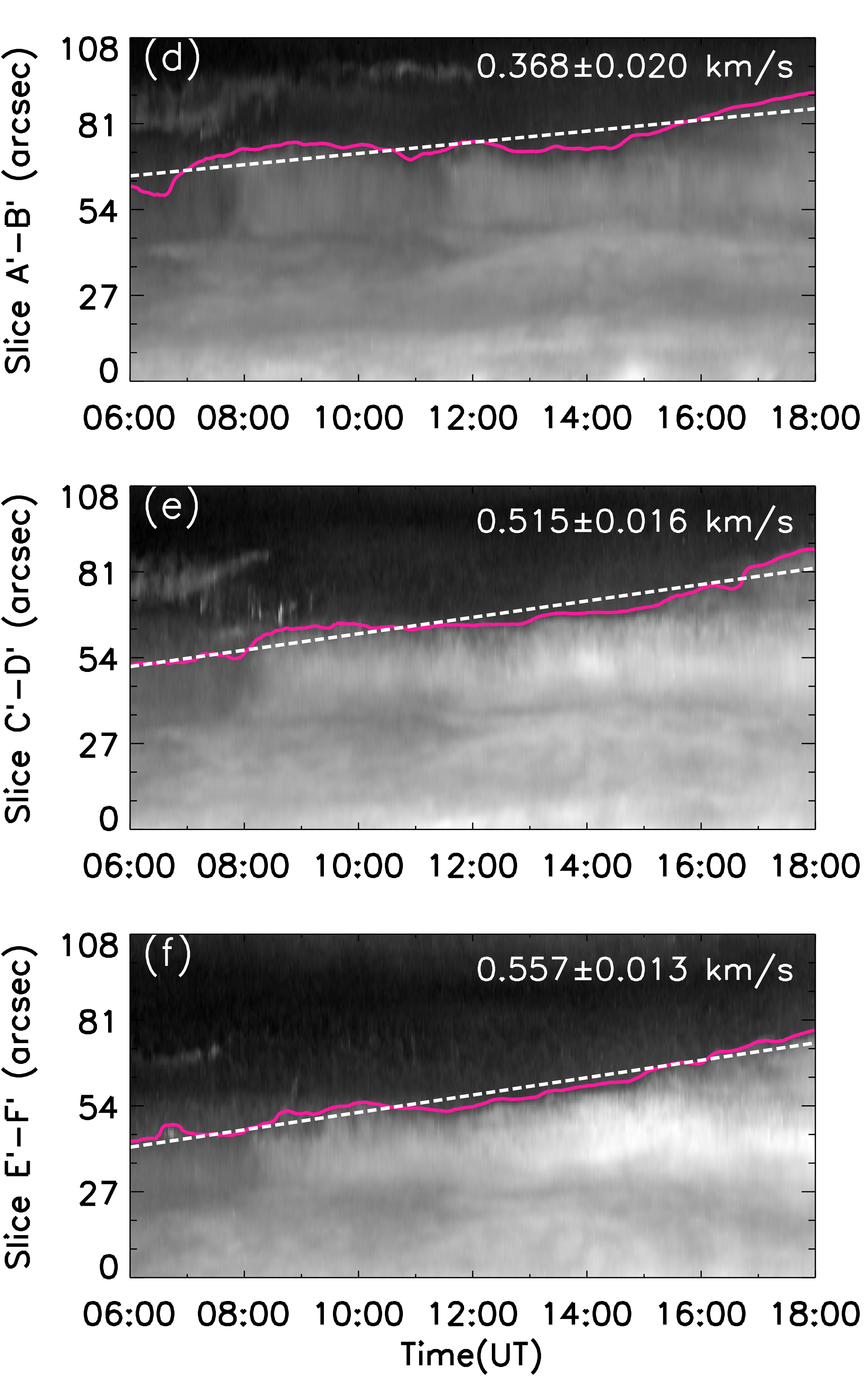}

\vspace*{0mm}
\caption{Space-time plots ($a$-c) along the slits ``A-B'', ``C-D'', and ``E-F'' presented in Figure 1(b6) and space-time plots (d-f) along the slits ``A$'$-B$'$'', ``C$'$-D$'$'', and ``E$'$-F$'$'' shown in Figure 1(b7). In panels ($a$-c), the magenta curves delineate the boundaries of the bright loops, and the dashed lines are linear fittings to the boundaries. In panels (d-f), the magenta curves delineate CHBs, and the dashed lines are linear fittings to the CHBs.}
\end{figure*}

\setcounter{figure}{2}
\begin{figure*}
\centering
\includegraphics[width=16cm]{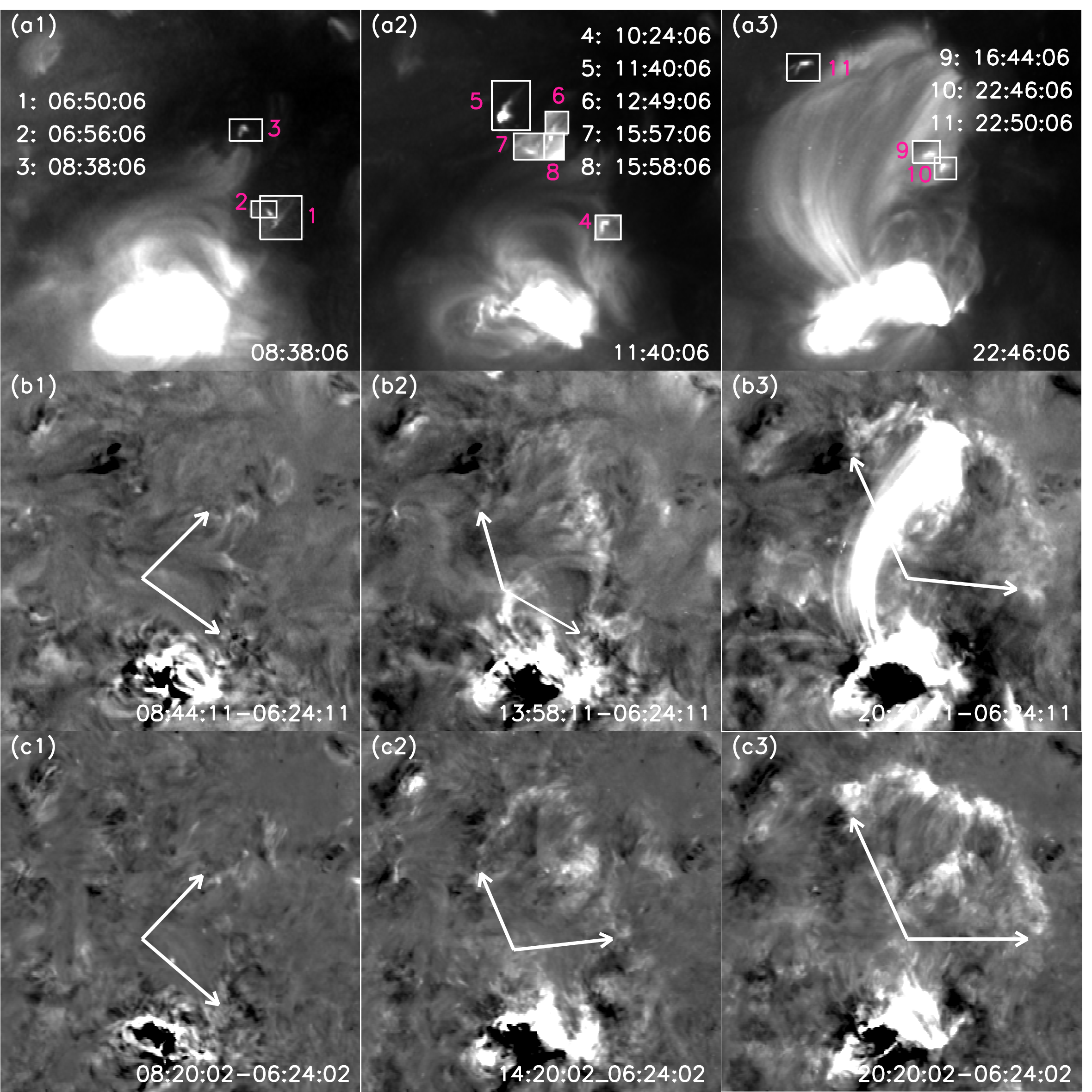}
 \caption{AIA 193 {\AA} direct images ($a$1-$a$3), AIA 171 {\AA} fixed-base difference images (b1-b3), and AIA 304 {\AA} fixed-base difference images (c1-c3). The sub-images outlined by white rectangles in panels ($a$1-$a$3) display the jets occurring at different times during the interaction. The white arrows in panels (b1-b3) and panels (c1-c3) indicate the arc-shaped brightening at the CHBs. The FOV is 160$^{''}$ $\times$ 160$^{''}$.}
\end{figure*}

\setcounter{figure}{3}
\begin{figure*}
\centering
\includegraphics[width=7.2cm]{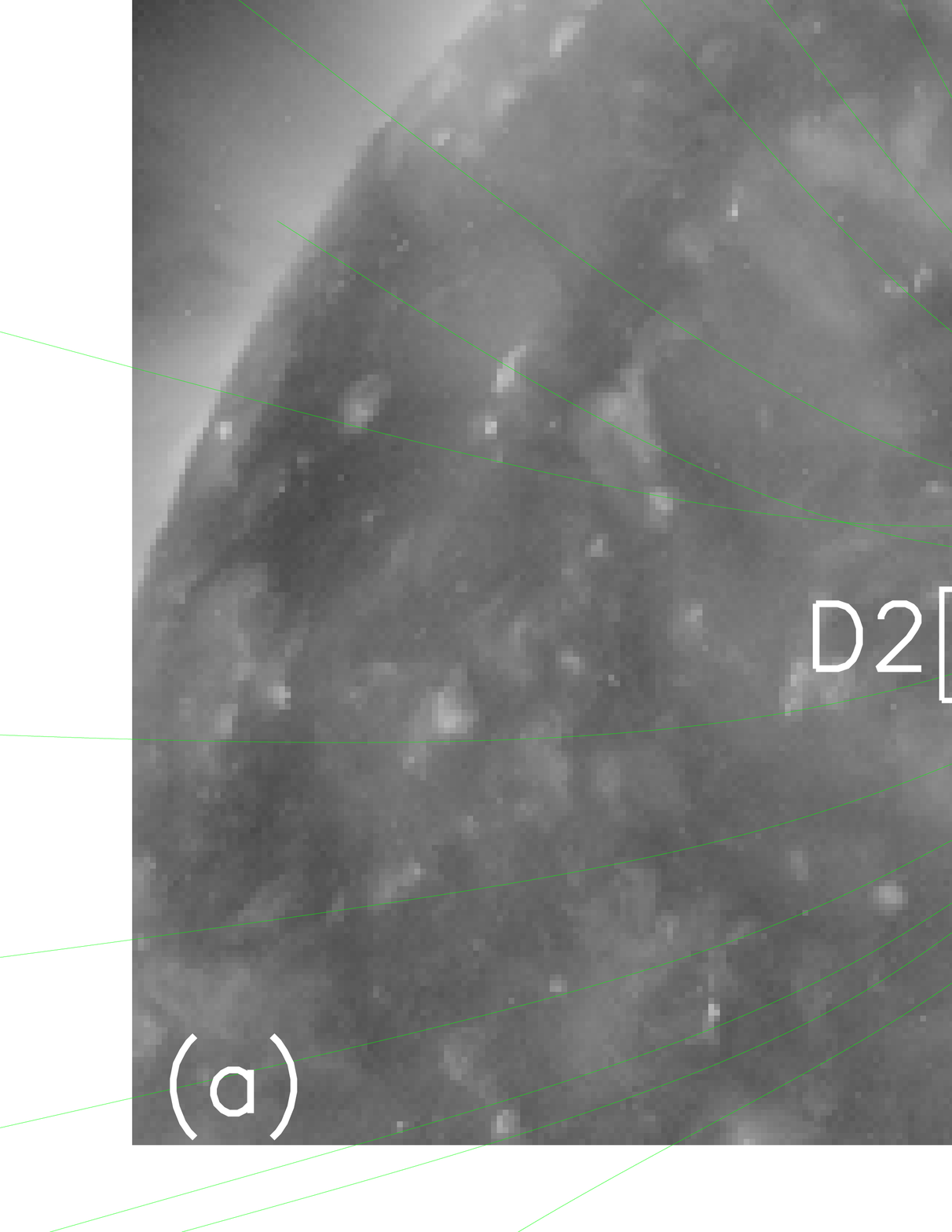}
\includegraphics[width=7.2cm]{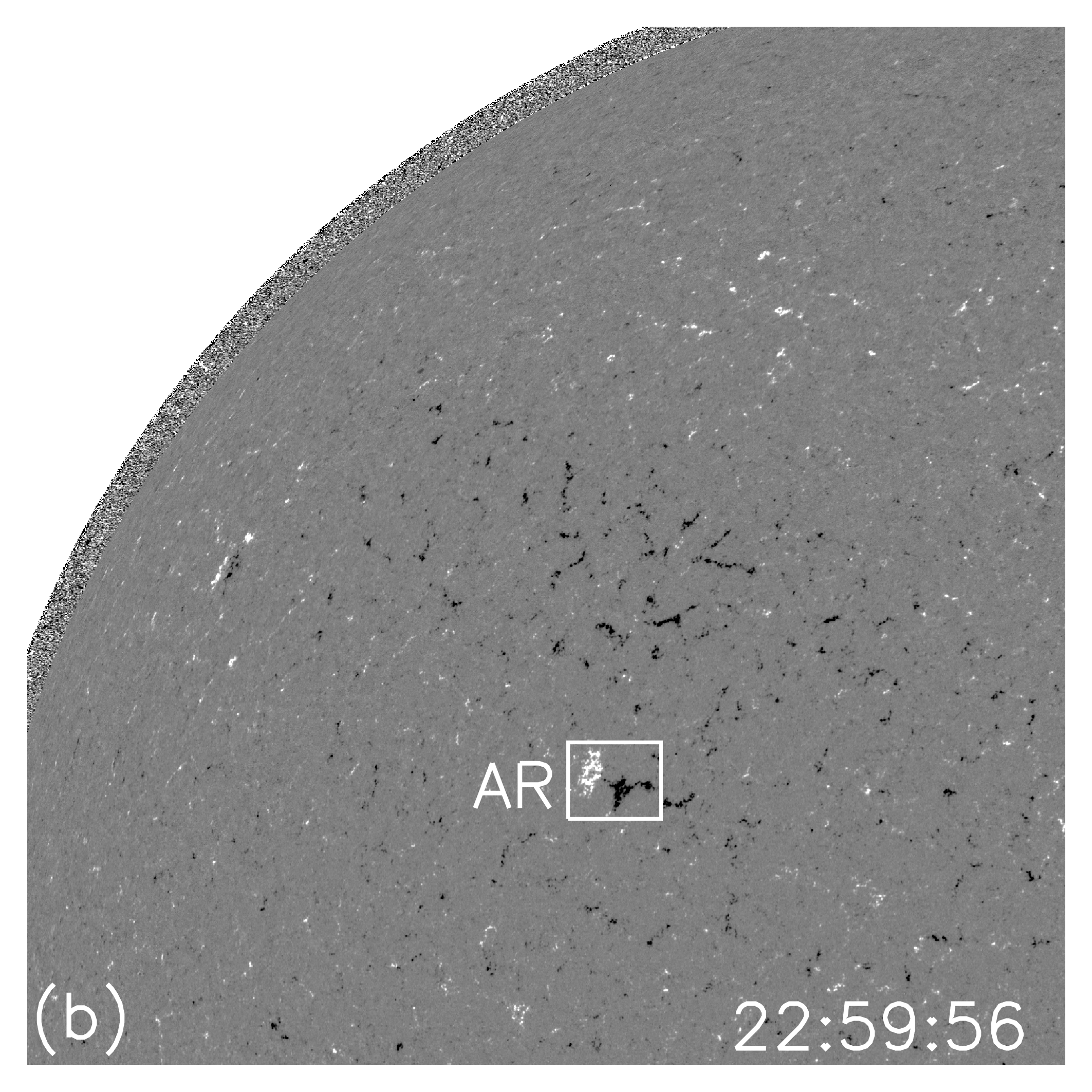}
\caption{AIA 193 {\AA} images with the superposition of extrapolated field lines ($a$) and HMI magnetogram (b). The yellow lines represent closed loops, while the green lines correspond to open field lines. The white rectangles in panel ($a$) indicate the dimming regions. The FOV is 900$^{''}$ $\times$ 900$^{''}$.}
\end{figure*}

\setcounter{figure}{4}
\begin{figure*}
\centering
\includegraphics[width=16cm]{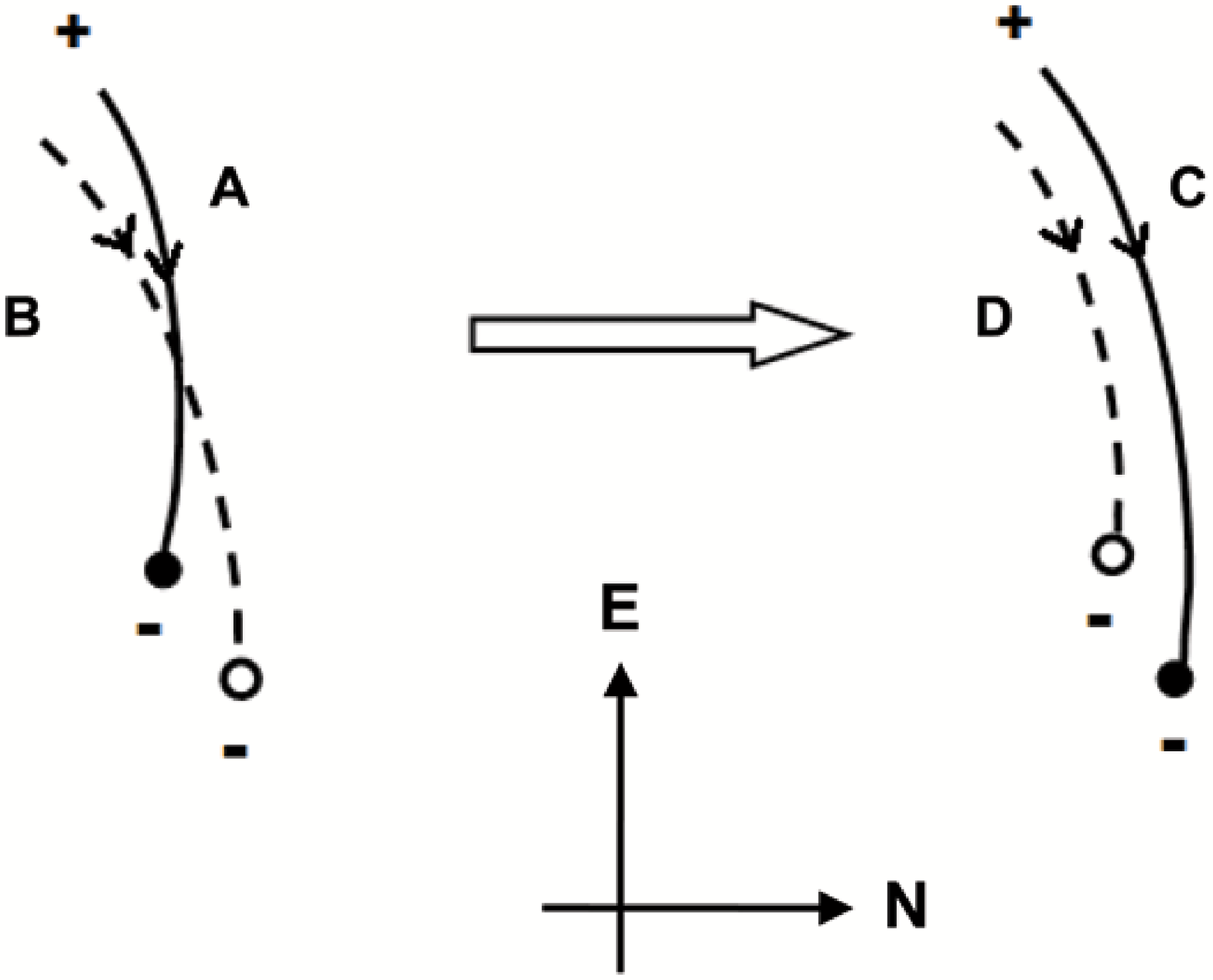}
\vspace*{-10mm}
\caption{Sketch of Y-type interchange reconnection. Solid (dashed) lines represent
closed (open) field lines. In the initial configuration, the closed loop of the AR
(A) is rooted just to the southeast of the open CH field line (B). Interchange reconnection at the apex of A  transports A northwestward to create new closed
loop (C) and B southeastward to form new open field line (D).}
\end{figure*}

\label{lastpage}


\begin{thebibliography}{}
\bibitem[1972]{}
Altschuler, M. D., Trotter, D. E., \& Orrall, F. Q. 1972, Sol. Phys., 26, 354
\bibitem[2006]{}
Attrill, G., Nakwacki, M. S., \& Harra, L. K., et al. 2006, Sol. Phys., 238, 117
\bibitem[2007]{}
Baker, D., van Driel-Gesztelyi, L., \& Attrill, G. D. R. 2007, AN., 328, 773
\bibitem[2002]{}
Crooker, N. U., Gosling, J. T., \& Kahler, S. W. 2002, J. Geophys. Res., 107, 1028
\bibitem[2005]{}
Fisk, L. A. 2005, ApJ, 626, 563
\bibitem[2001]{}
Fisk, L. A., \& Schwadron, N. A. 2001, ApJ, 560, 425
\bibitem[1999]{}
Fisk, L. A., Zurbuchen, T. H., \& Schwadron, N. A. 1999, ApJ, 521, 868
\bibitem[2007]{}
Jiang, Y. C., Yang, L. H., Li, K. J., et al. 2007, ApJ, 667, L105
\bibitem[2002]{}
Kahler, S. W., \& Hudson, H. S. 2002, ApJ, 574, 467
\bibitem[1990]{}
Kahler, S. W., \& Moses, D. 1990, ApJ, 362, 728
\bibitem[1973]{}
Krieger, A. S., Timothy, A. F., \& Roelof, E. C. 1973, Sol. Phys., 29, 505
\bibitem[2012]{}
Lemen, J. R., Title, A. M., \& Akin, D. J., et al. 2012, Sol. Phys., 275, 17
\bibitem[2004]{}
Madjarska, M. S., Doyle, J. G., \& van Driel-Gesztelyi, L. 2004, ApJ, 603, L57
\bibitem[2009]{}
Madjarska, M. S., \& Wiegelmann, T. 2009, A\&A, 503, 991
\bibitem[1972]{}
Munro, R. H., \& Withbroe, G. L. 1972, ApJ, 176, 511
\bibitem[1978]{}
Nolte, J. T., Davis, J. M., \& Gerassimenko, M., et al. 1978, Sol. Phys., 60, 143
\bibitem[2001]{}
Scherrer, P. H., Bogart, R. S., \& Bush, R. I., et al. 1995, Sol. Phys., 162, 129
\bibitem[2012]{}
Schou, J., Scherrer, P. H., \& Bush, R. I., et al. 2012, Sol. Phys., 275, 229
\bibitem[2003]{}
Schrijver, C. J., \& DeRosa, M. L. 2003, Sol. Phys., 212, 165
\bibitem[2010]{}
Subramanian, S., Madjarska, M. S., \& Doyle, J. G. 2010, A\&A, 516, 50
\bibitem[1975]{}
Timothy, A. F., Krieger, A. S., \& Vaiana, G. S. 1975, Sol. Phys., 42, 135
\bibitem[2005]{}
Tu, C. Y., Zhou, C., \& Marsch, E., et al. 2005, Science, 308, 519
\bibitem[1976]{}
Vaiana, G. S., Zombeck, M., Krieger, A. S., \& Timothy, A. F. 1976, Ap\&SS, 39, 75
\bibitem[1996]{}
Wang, Y. M., Hawley, S. H., \& Sheeley, N. R., Jr. 1996, Science, 271, 464
\bibitem[2004]{}
Wang, Y. -M., \& Sheeley, N. R., Jr. 2004, ApJ, 612, 1196
\bibitem[2011]{}
Yang, S. H., Zhang, J., Li, T., \& Liu, Y. 2011, ApJ, 732, L7
\bibitem[2010]{}
Yokoyama, M., \& Masuda, S. 2010, Sol. Phys., 263, 135
\bibitem[1995]{}
Yokoyama, T., \& Shibata, K. 1995, Nature, 375, 42
\end{thebibliography}
\end{document}